\documentclass[conference]{IEEEtran}
\IEEEoverridecommandlockouts
\usepackage{multirow}
\usepackage{cite}
\usepackage{amsmath,amssymb,amsfonts}
\usepackage{algorithmic}
\usepackage{graphicx}
\usepackage{textcomp}
\usepackage{xcolor}
\usepackage{booktabs}
\def\BibTeX{{\rm B\kern-.05em{\sc i\kern-.025em b}\kern-.08em
    T\kern-.1667em\lower.7ex\hbox{E}\kern-.125emX}}
\begin{document}
\title{ Multi-Source Spatial Knowledge Understanding for Immersive Visual Text-to-Speech
}

\author{
\IEEEauthorblockN{
Shuwei He}
\IEEEauthorblockA{\textit{Inner Mongolia University} \\
Hohhot, China \\
shuwei\_he@163.com}
\and

\IEEEauthorblockN{
Rui Liu* \thanks{* Corrposending Author.}}
\IEEEauthorblockA{\textit{Inner Mongolia University} \\
Hohhot, China \\
liurui\_imu@163.com}

}

\maketitle

\begin{abstract}
Visual Text-to-Speech (VTTS) aims to take the environmental image as the prompt to synthesize reverberant speech for the spoken content. 
Previous works focus on the RGB modality for global environmental modeling, overlooking the potential of multi-source spatial knowledge like depth, speaker position, and environmental semantics. 
To address these issues, we propose a novel multi-source spatial knowledge understanding scheme for immersive VTTS, termed MS$^2$KU-VTTS. 
Specifically, we first prioritize RGB image as the dominant source and consider depth image, speaker position knowledge from object detection, and Gemini-generated semantic captions as supplementary sources. 
Afterwards, we propose a serial interaction mechanism to effectively integrate both dominant and supplementary sources.
The resulting multi-source knowledge is dynamically integrated based on the respective contributions of each source.
This enriched interaction and integration of multi-source spatial knowledge guides the speech generation model, enhancing the immersive speech experience.
Experimental results demonstrate that the MS$^2$KU-VTTS surpasses existing baselines in generating immersive speech. 
Demos and code are available at: {https://github.com/AI-S2-Lab/MS2KU-VTTS}.
\end{abstract}


\begin{figure*}[ht]

    \centering    
    \includegraphics[width=1\linewidth]{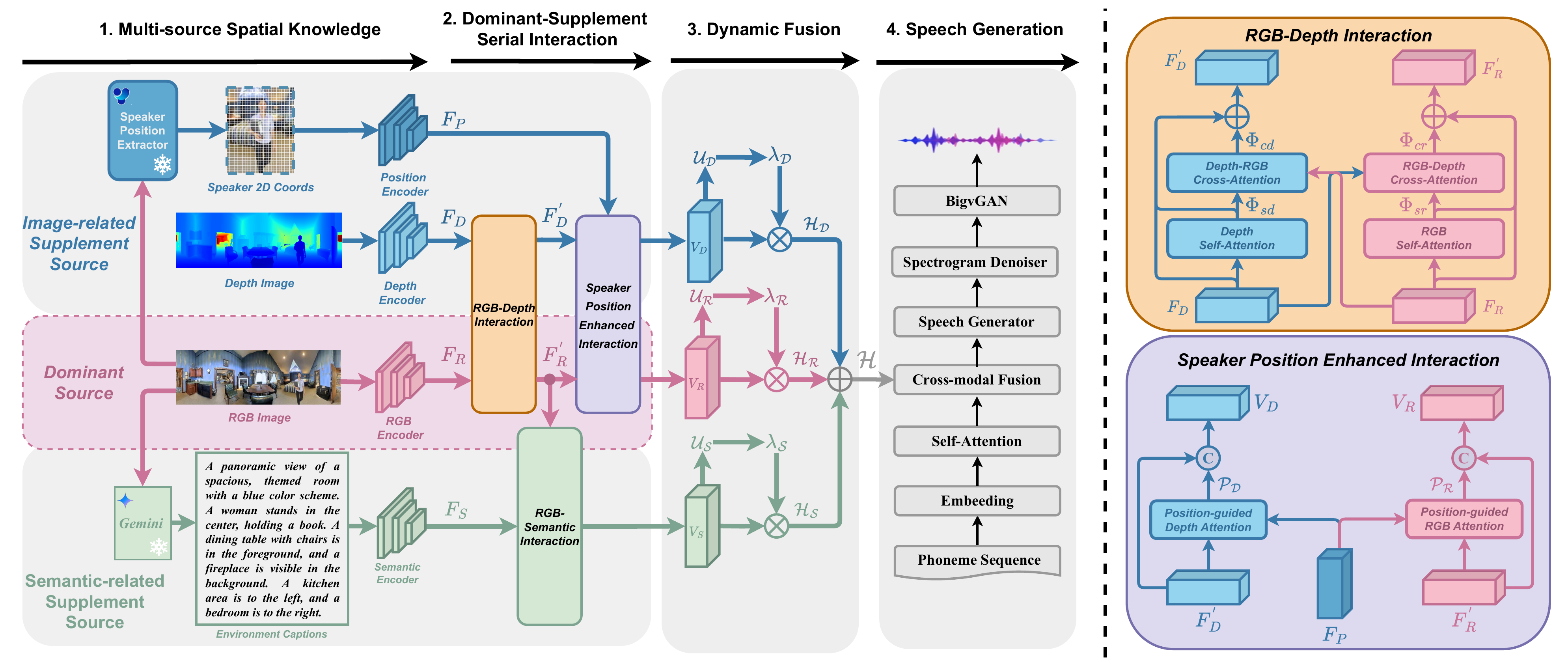}
    \vspace{-8mm}
    \caption{The overall architecture of MS$^2$KU-VTTS.}
    \label{fig:model}
    \vspace{-5mm}
\end{figure*}

\section{Introduction}

Visual Text-to-Speech (VTTS) aims to generate reverberant speech that matches spoken content based on the spatial environmental image. 
With advancements in human-computer interaction, VTTS has become an integral part of intelligent systems, playing a vital role in fields such as augmented reality and virtual reality.

Unlike acoustic matching tasks, which focus on adapting input speech to match a reference environment \cite{vam, svam, audioldm, diffrent}, VTTS aims to synthesize speech that incorporates the environmental characteristics of that environment based on given textual content \cite{m2se}. For example, VoiceLDM \cite{voiceldm} utilizes a pre-trained CLAP model \cite{clap} to transform text or audio descriptions into environmental feature vectors, which control the reverberation characteristics of the generated audio. In addition, Environment-aware TTS \cite{eatts} designs an environment embedding extractor that learns environmental features from reference speech by optimizing embedding distances to distinguish between different environments. In a recent study, ViT-TTS \cite{vit-tts} introduces a transformer-based visual-text encoder for extracting global spatial knowledge from an RGB image. Building on these advancements, this paper focuses on employing visual information as a guide to synthesize reverberant audio for the target environment.

However, prior research in VTTS predominantly focuses on RGB images as the sole source of spatial knowledge. Therefore, integrating multi-source spatial data—including depth images, speaker position data, and spatial semantics—could significantly enhance spatial understanding. For example, FEW-SHOTRIR \cite{avlea} employs depth images to enhance its understanding of environmental acoustics, significantly improving its ability to predict room impulse responses by incorporating key spatial contexts and geometric details for precise audio simulations in unfamiliar settings. Moreover, speaker position plays a crucial role in shaping the perception of reverberation \cite{lavd}. For instance, ViGAS \cite{NovelView} utilizes speaker position data to refine the accuracy of spatial audio synthesis by aligning visual cues with auditory outputs. This approach enhances the realism and spatial coherence of audio in novel-view acoustic synthesis tasks. Additionally, environment captions provide detailed spatial knowledge about environmental layouts and dimensions, enhancing situational adaptability \cite{voiceldm}. For instance, AST-LDM \cite{Speak} leverages textual descriptions of target environments as reference prompts to guide the acoustic scene transfer process, enhancing the model’s versatility and applicability in generating immersive auditory experiences.

To address the above challenges, we propose a novel multi-source spatial knowledge understanding scheme for immersive VTTS, termed \textbf{MS$^2$KU-VTTS}. Specifically, to expand the dimensions of environmental perception, we first prioritize the RGB image as \textbf{the dominant source} and consider depth image, speaker position knowledge from object detection as \textbf{the image-related supplement source}, and semantic captions from Gemini Pro Vision \cite{gemini} for image understanding as \textbf{the semantic-related supplementary source}.Afterwards, to deeply engage with both dominant and supplementary sources, we propose a Dominant-Supplement Serial Interaction mechanism to process features extracted from various knowledge sources. It is through this sequential approach that not only is the model's perception of spatial relationships deepened, but also, via complex interactions between dominant and supplementary sources, precise modeling of environmental reverberations is facilitated. In the end, for comprehensive environmental understanding, the resulting multi-source knowledge is \textit{dynamically} integrated based on each source's contribution. This enriched interaction and integration of multi-source spatial knowledge guides the speech generation model, enhancing the immersive speech experience. We conducted extensive experiments to evaluate the effectiveness of MS$^2$KU-VTTS. Both subjective and objective evaluations demonstrate that our model outperforms state-of-the-art baseline systems in terms of naturalness and perceptual quality.

\section{Method}

\subsection{Model Overview}

As illustrated in the pipeline in Fig.\ref{fig:model}, the proposed MS$^2$KU-VTTS architecture consists of four components: 1) \textit{Multi-source Spatial Knowledge}, constructed from both dominant and supplementary sources; 2) \textit{Dominant-Supplement Serial Interaction} aggregates intra- and inter-modal information through multiple mechanisms. Specifically, \textit{RGB-Depth Interaction} enables comprehensive spatial understanding, \textit{Speaker Position Enhanced Interaction} employs positional data to model environmental acoustics, and \textit{RGB-Semantic Interaction} extracts spatial semantics from Gemini-generated captions; 3) \textit{Dynamic Fusion}, wherein the multi-source knowledge is dynamically fused based on the contributions of each source; and 4) \textit{Speech Generation}, which uses the integrated information to guide the synthesis of reverberant speech.

\subsection{Multi-source Spatial Knowledge}
The dominant source, encompassing the RGB image, is supplemented by the depth image, speaker position from object detection, and semantic captions.

\subsubsection{Dominant Source - RGB}
To learn the global knowledge in the RGB image, such as the material and color of objects, features are derived through a pre-trained ResNet18 \cite{resnet18}, denoted as $F_R \in \mathcal{R}^{1 \times D}$, where $D$ represents the feature dimension.

\subsubsection{Image-related Supplement Source - Depth and Speaker Position}
To capture the global knowledge within the depth image, such as object arrangement, depth features are also extracted using a ResNet18, denoted as $F_D \in \mathcal{R}^{1 \times D}$. To acquire fine-grained spatial knowledge, this work integrates the positional information of the speaker as additional guidance, inspired by previous studies \cite{lavss, transformer, nerf, NovelView}. Specifically, we first detect all recognizable objects within the RGB image to obtain their corresponding anchor box coordinates. Subsequently, those coordinates categorized as ``human" are selected and transformed into two-dimensional pixel coordinates $(x,y)$, which are normalized to the range $[0, 1]$ based on the RGB image width and height, respectively. These coordinates are then mapped into a higher-dimensional space. The mathematical representation is as follows:
\begin{equation} 
\begin{split}
\phi(x, y) = \Big( & \sin(2^k \pi x), \cos(2^k \pi x), \\
& \sin(2^k \pi y), \cos(2^k \pi y) \Big)_{k=0}^{D-1}
\end{split}
\end{equation}

where $\phi(\cdot)$ denotes the high-dimensional embedding of the position of speaker, employing harmonic terms up to a dimension of $D$. The embeddings are subsequently processed by adaptive max pooling followed by a multi-layer perceptron to extract positional features, denoted as $F_P \in \mathcal{R}^{1 \times D}$.

\subsubsection{Semantic-related Supplement Source - Spatial Image Captions}
To efficiently fuse semantic information within the RGB image, Gemini Pro Vision \cite{gemini} is employed to transform complex visual information into structured captions, which are subsequently processed by BERT \cite{bert} to extract spatial semantic features, denoted as $F_S$, where $F_S \in \mathcal{R}^{1 \times D}$.

\subsection{Dominant-Supplement Serial Interaction}
As shown in the middle panel of Fig.\ref{fig:model}, the Dominant-Supplement Serial Interaction (D-SSI) comprises three components: 1) RGB-Depth Interaction, which aggregates knowledge from dominant and image-related sources; 2) Speaker Position Enhanced Interaction, which models the relationship between source position and environmental acoustics; and 3) RGB-Semantic Interaction, which learns rich spatial semantics.

\subsubsection{RGB-Depth Interaction}
The RGB-Depth Interaction, inspired by LGPM \cite{pstm}, is introduced to achieve a more comprehensive representation of RGB and Depth features. It first refines the feature representations of each source through self-attention mechanisms, while simultaneously enabling cross-modal information fusion via cross-attention. After that, these representations are aggregated to update the RGB and Depth features $F_R^{\prime}$ and $F_D^{\prime}$, which are formulated as:
\begin{equation} 
F_D^{\prime} = F_R + \Phi_{sr}(F_R, F_R, F_R) + \Phi_{cr}(F_R, F_D, F_D), 
\end{equation} 
\begin{equation} 
F_R^{\prime} = F_D + \Phi_{sd}(F_D, F_D, F_D) + \Phi_{cd}(F_D, F_R, F_R),
\end{equation} where $\Phi_{s*}(\cdot)$ and $\Phi_{c*}(\cdot)$ denote the self-attention and cross-modal attention functions, respectively.

\subsubsection{Speaker Position Enhanced Interaction}
To capture detailed spatial knowledge    within a visual scene, we propose an attention-based Speaker Position Enhanced Interaction. Specifically, the position-related visual features, denoted as $\mathcal{P}_{R/D}$, computed as:
\begin{equation} 
\mathcal{P}_{R/D} = F_{R/D}^{\prime} \cdot \varphi((F_P)^T \cdot F_{R/D}^{\prime}), 
\end{equation} 
where $\varphi(\cdot)$ denotes the softmax function and $(\cdot)^T$ represents the transpose operator. Subsequently, the position-related and visual features are combined through concatenation: 
\begin{equation} 
\mathcal{V}_{R/D} = \varphi(FC(Concat[F_{R/D}^{\prime}, \mathcal{P}_{R/D}])), 
\end{equation} where $FC$ represents fully connected layers. 

\subsubsection{RGB-Semantic Interaction}
To capture spatial semantics from environmental captions, we employ Spatial Semantic Attention by applying a multi-head attention mechanism: 
\begin{equation} 
\mathcal{V}_S = \text{MultiHead}(F_S, F_R^{\prime}, F_R^{\prime}), 
\end{equation} where $V_S \in \mathbb{R}^{1 \times D}$ is derived by refining $F_S$.

\subsection{Dynamic Fusion}
To effectively aggregate multi-source spatial knowledge, we employ a dynamic weighting approach inspired by MLA \cite{mla}, which adjusts the contribution of each type of spatial knowledge based on its complexity, quantified by entropy. The underlying assumption is that higher entropy indicates greater uncertainty or difficulty in understanding, which reduces the relative importance of that knowledge type in the fusion process. The entropy for each knowledge type, denoted as $u_i$, is computed as:
\begin{equation} 
u_i = -\sum_{j=1}^{D} p_{(i, j)} \log p_{(i, j)}, 
\end{equation} where $p_{(i, j)} = \text{Softmax}(V_i)$ represents the probability distribution over the features of a given type $i$. The relative importance $\lambda_i$ for each knowledge type is then determined by: 
\begin{equation} 
\lambda_i = \frac{\exp(u_{max} - u_i)}{\sum_{k=1}^{M} \exp(u_{max} - u_k)}, 
\end{equation} where $u_{max} = \max(u_1, ..., u_M)$, with $M$ denoting the total number of spatial knowledge types. The representation of integrated multi-source knowledge as $\mathcal{H}$ is defined as follows: 
\begin{equation} 
\mathcal{H} = \lambda_R \cdot \mathcal{H}_R + \lambda_D \cdot \mathcal{H}_D + \lambda_S \cdot \mathcal{H}_S.
\end{equation}

\subsection{Speech Generation}
As depicted in Fig. \ref{fig:model}, our TTS system utilizes the ViT-TTS architecture. Initially, phoneme embeddings and visual features are converted into hidden sequences. These sequences are then adjusted by a speech generator to synchronize with speech frames. A spectrogram denoiser refines the adjusted hidden states into mel-spectrograms. For a comprehensive methodology, refer to ViT-TTS \cite{vit-tts}.

\section{Experiments and Results}
\vspace{-2mm}
\begin{table*}[t!]
\caption{Comparison of MS$^2$KU-VTTS and baselines on the SoundSpaces-Speech dataset for Seen and Unseen scenarios, including subjective (95\% confidence intervals) and objective metrics. MS$^2$KU-VTTS significantly outperforms baselines ($p < 0.001$).}
\vspace{-4mm}
\begin{center}
 
\resizebox{0.75\linewidth}{!}{
\begin{tabular}{@{}l|ccc|ccc@{}}
\toprule
\multirow{2}{*}{\textbf{System}}              & \multicolumn{3}{c|}{\textbf{Test-Unseen}}                                                                                                & \multicolumn{3}{c}{\textbf{Test-Seen}}                                                                                                   \\
                                              & \textbf{MOS} ($\uparrow$) & \textbf{RTE} ($\downarrow$) & \textbf{MCD} ($\downarrow$) & \textbf{MOS} ($\uparrow$) & \textbf{RTE} ($\downarrow$) & \textbf{MCD} ($\downarrow$) \\ \midrule
GT                                            & 4.353 $\pm $ 0.023                         & /                                            & /                                            & 4.348 $\pm $ 0.022                         & /                                            & /                                            \\
GT(voc.)                                      & 4.149 $\pm $ 0.027                         & 0.0080                                       & 1.4600                                       & 4.149 $\pm $ 0.023                         & 0.0060                                       & 1.4600                                       \\ \midrule
ProDiff  \cite{huang2022prodiff}       & 3.550 $\pm $ 0.023                         & 0.1341                                       & 4.7689                                       & 3.647 $\pm $ 0.023                         & 0.1243                                       & 4.6711                                       \\
DiffSpeech  \cite{diffsinger} & 3.649 $\pm $ 0.022                         & 0.1193                                       & 4.7923                                       & 3.675 $\pm $ 0.011                         & 0.1034                                       & 4.6630                                       \\
VoiceLDM  \cite{voiceldm}     & 3.702 $\pm $ 0.020                         & 0.0825                                       & 4.8952                                       & 3.702 $\pm $ 0.025                         & 0.0714                                       & 4.6572                                       \\
ViT-TTS  \cite{vit-tts}       & 3.700 $\pm $ 0.025                         & 0.0759                                       & 4.5933                                       & 3.804 $\pm $ 0.022                         & 0.0677                                       & 4.5535                                       \\ \midrule
\textbf{MS$^2$KU-VTTS}                        & \textbf{3.875 $\pm $ 0.011}                & \textbf{0.0745}                              & \textbf{4.5544}                              & \textbf{3.947 $\pm $ 0.022}                & \textbf{0.0668}                              & \textbf{4.5175}                              \\ \bottomrule
\end{tabular}}
\label{table:baseline}
\end{center}
\vspace{-7mm}
\end{table*}
\subsection{Datasets} 
Utilizing the SoundSpaces-Speech dataset \cite{lavd}, we follow the approach outlined in \cite{vam, vit-tts}, which involves removing out-of-view samples and partitioning the data into ``test-unseen"—featuring room acoustics from novel scenes—and ``test-seen"—comprising scenes encountered during training. The dataset encompasses 28,853 training, 1,441 validation, and 1,489 testing samples, each containing clean text, reverberant audio, and panoramic RGB-D images. Text sequences are converted into phonemes using an open-source tool\footnote{https://github.com/Kyubyong/g2p}. Consistent with \cite{ren2021fastspeech, huang2022prodiff, ECSS, catts}, spectrograms are extracted via FFT with parameters: size 1024, hop size 256, and window size 1024 samples; these are then transformed into 80-bin mel-spectrograms. Fundamental frequency is extracted from raw waveforms using Parselmouth\footnote{https://github.com/YannickJadoul/Parselmouth}.

\subsection{Implementation Details}
The feature dimension is 512. The speaker position is detected using YOLOv8 \footnote{https://github.com/ultralytics/ultralytics}. Gemini Pro Vision analyzes spatial semantics by prompting:  \textquotedblleft Observe this RGB panoramic image and briefly describe its contents, including specific objects or people and their locations. Detail the spatial relationships among them, such as which object is to the left, right, above, or below others. Please focus on key information only.\textquotedblright The phoneme set includes 74 distinct phonemes, aligned with ViT-TTS \cite{vit-tts} speech generation parameters. Training involves two phases: 1) a pre-training stage for the encoder following ViT-TTS protocols over 120k steps until convergence; 2) a main training stage on an NVIDIA A800 GPU, processing 48 sentences per batch for 160k steps. For inference, BigVGAN \cite{bigvgan} uniformly converts mel-spectrograms into waveforms.

\subsection{Evaluation Metrics}
We evaluate waveform quality using both objective metrics and subjective assessments, aligning with previous studies \cite{diffsinger, huang2022prodiff}. For objective analysis, we randomly select 50 samples from the test set. Our evaluation metrics include: 1) Perceptual quality, assessed by human listeners and quantified by the Mean Opinion Score (MOS), which rates audio quality on a scale of 1 to 5; 2) Room acoustics accuracy, measured by RT60 Error (RTE), which compares the reverberation time of predicted and target waveforms; 3) Mel Cepstral Distortion (MCD) \cite{mcd}, which measures the spectral distance between synthesized and reference mel-spectra.

\subsection{Baseline Models}
ProDiff \cite{huang2022prodiff}, DiffSpeech \cite{diffsinger}, and VoiceLDM \cite{voiceldm} are diffusion-based TTS models that accept speech text as input, with the latter also requiring environmental captions. In our implementation, captions focus on specific components and spatial relations, rather than types of environments. ViT-TTS \cite{vit-tts} represents a VTTS system.

\subsection{Main Results}

As presented in Table \ref{table:baseline}, MS$^2$KU-VTTS demonstrates superior performance across all evaluation metrics, achieving the best RTE (0.0745), MCD (4.5544), and MOS (3.875 $\pm$ 0.011) in unseen environments. Statistical analysis indicates that MS$^2$KU-VTTS significantly outperforms the baselines with a p-value less than 0.001. Although efficacy slightly decreases on the test-unseen set compared to seen scenarios—likely due to unfamiliar training conditions—the model still surpasses DiffSpeech \cite{diffsinger} and ProDiff \cite{huang2022prodiff} across all metrics, particularly in RTE, underscoring the limitations of traditional TTS models with spatial environmental data. Our multi-source approach effectively captures comprehensive spatial knowledge, conferring significant advantages. The model surpasses VoiceLDM \cite{voiceldm} due to our advanced Gemini-based image understanding, which outperforms VoiceLDM’s simpler environmental prompts. While ViT-TTS \cite{vit-tts} exceeds other baselines, it underperforms relative to our model in both environments, confirming the efficacy of our approach in generating environment-aligned reverberant speech.

\begin{table}[htbp]
\caption{Ablation study results. Depth, Position, RGB, Semantics, R-DI, R-SI, and DF represent different knowledge components and interactions. Full MS$^2$KU-VTTS results are in Table \ref{table:baseline}.}
\begin{center}
\resizebox{0.85\linewidth}{!}{
\begin{tabular}{@{}l|ccc@{}}
\toprule
\textbf{System}       & \textbf{MOS} ($\uparrow$) & \textbf{RTE} ($\downarrow$) & \textbf{MCD} ($\downarrow$) \\ \midrule
GT(voc.)              & 4.149 $\pm $ 0.027                         & 0.0080                                       & 1.4600                                       \\ \midrule
w/o Depth             & 3.747 $\pm $ 0.020                         & 0.1091                                       & 4.6593                                       \\
w/o Position          & 3.754 $\pm $ 0.019                         & 0.1051                                       & 4.8008                                       \\
w/o RGB               & 3.728 $\pm $ 0.011                         & 0.0983                                       & 4.8898                                       \\
w/o Semantics         & 3.745 $\pm $ 0.023                         & 0.1048                                       & 4.8559                                       \\
w/o R-DI              & 3.753 $\pm $ 0.022                         & 0.0985                                       & 4.8742                                       \\
w/o R-SI              & 3.796 $\pm $ 0.019                         & 0.0962                                       & 4.8508                                       \\
w/o DF                & 3.672 $\pm $ 0.035                         & 0.1027                                       & 4.9972                                       \\ \midrule
\textbf{MS$^2$KU-VTTS} & \textbf{3.875 $\pm $ 0.011}                & \textbf{0.0745}                              & \textbf{4.5544}                              \\ \bottomrule
\end{tabular}}
\label{table:wo}
\end{center}
\vspace{-7mm}
\end{table}

\subsection{Ablation Results}

To assess the individual contributions of key components within our model on the Test-Unseen set, we conducted ablation studies by systematically removing these elements. The results, encompassing both subjective and objective metrics, are displayed in Table \ref{table:wo}. Removing various sources of knowledge from the Multi-source Spatial Knowledge Extraction led to declines in objective metrics and subjective MOS scores, underscoring the efficacy of our multi-source approach in enhancing reverberation expressiveness. For instance, the removal of Position knowledge resulted in a MOS decrease of 0.121, while RTE and MCD increased by 0.0306 and 0.2464, respectively. To evaluate the Dominant-Supplement Serial Interaction, the removal of various interaction modules led to reductions in all measured scores, highlighting the adequacy of our interactions across spatial knowledge types. Lastly, substituting Dynamic Fusion with simple concatenation diminished performance across all metrics, confirming that Dynamic Fusion fosters a more comprehensive understanding of multi-source knowledge.

\section{CONCLUSION AND FUTURE WORK}
In this paper, we introduce a novel multi-source spatial knowledge understanding scheme, termed MS$^2$KU-VTTS, capable of generating immersive, environment-matched reverberant speech. The proposed Dominant-Supplement Serial Interaction and Dynamic Fusion ensure precise modeling of overall environmental reverberation and a comprehensive understanding of multi-source knowledge, respectively. Experimental results affirm the superiority of MS$^2$KU-VTTS over contemporary VTTS systems. In the future, we will focus on optimizing computational efficiency and enhancing the model’s adaptability to diverse and previously unencountered spatial environments.

\clearpage

\section{Acknowledgments}

This work was funded by the Young Scientists Fund (No. 62206136) and the General Program (No. 62476146) of the National Natural Science Foundation
of China, the ``Inner Mongolia Science and Technology Achievement Transfer and Transformation Demonstration Zone, University Collaborative Innovation Base, and University Entrepreneurship Training Base'' Construction Project (Supercomputing Power Project) (No.21300-231510).

\bibliographystyle{IEEEtran}
\bibliography{refs}
\end{document}